\documentclass{PoS}

\title{On the Validity of the Effective Field Theory for Dark Matter Searches at the LHC}

\ShortTitle{EFT for Dark Matter Searched at LHC}

\author{\speaker{Giorgio Busoni}\\
        SISSA and INFN, Sezione di Trieste, via Bonomea 265, I-34136 Trieste, Italy\\
        E-mail: \email{giorgio.busoni@sissa.it}}

\author{Andrea De Simone\\
        SISSA and INFN, Sezione di Trieste, via Bonomea 265, I-34136 Trieste, Italy\\
        E-mail: \email{andrea.desimone@sissa.it}}
\author{Johanna Gramling\\
        Section de Physique, Universit\'e de Gen\`eve,\\
24 quai E. Ansermet, CH-1211 Geneva, Switzerland\\
        E-mail: \email{johanna.gramling@unige.ch}}
\author{Enrico Morgante\\
        Section de Physique, Universit\'e de Gen\`eve,\\
24 quai E. Ansermet, CH-1211 Geneva, Switzerland\\
        E-mail: \email{enrico.morgante@unige.ch}}
\author{Antonio Riotto\\
        Section de Physique, Universit\'e de Gen\`eve,\\
24 quai E. Ansermet, CH-1211 Geneva, Switzerland\\
        E-mail: \email{antonio.riotto@unige.ch}}

\abstract{We generalize in several directions our  recent analysis of the limitations to the use of the effective field theory approach to study dark matter at the LHC. Firstly, we study the full list of operators connecting  fermion DM to quarks and gluons,  corresponding to integrating out a heavy mediator
in the $s$-channel; secondly,  we provide analytical results for the validity of the EFT description for both  $\sqrt{s}=8$ {\rm TeV} and $14$ {\rm TeV}; thirdly,
we make use of a  MonteCarlo event generator approach to assess the validity of our analytical conclusions. We apply our results to revisit the current collider  bounds on the ultraviolet cut-off scale of the effective field
theory and show that these bounds are weakened once
 the validity conditions of the  effective field theory are imposed.}

\FullConference{XXII. International Workshop on Deep-Inelastic Scattering and Related Subjects,\\
		28 April - 2 May 2014\\
		Warsaw, Poland}

\usepackage{amsmath}
\usepackage{amsfonts}
\usepackage{amssymb}
\usepackage{graphicx}
\usepackage{mathrsfs}
\usepackage{latexsym}
\usepackage{cite}
\usepackage{slashed, cancel}

\makeatother   
\newcommand{\GeV}{{\rm \,GeV}}
\newcommand{\TeV}{{\rm TeV}}

\def\de{\textrm{d}}

 \def\be   {\begin{equation}}   \def\ee   {\end{equation}}
 \def\ba   {\begin{array}}      \def\ea   {\end{array}}
 \def\bea  {\begin{eqnarray}}   \def\eea  {\end{eqnarray}}
 \def\bean {\begin{eqnarray*}}  \def\eean {\end{eqnarray*}}

\begin{document}

\section{Introduction}
While there are many  cosmological and astrophysical evidences that our universe contains
a sizable amount of dark Matter (DM),  i.e.~a component which clusters at small scales, its nature is still a mystery. 
Currently, there are several ways to search for such DM
candidates. DM particles (if they are light enough) might reveal themselves in particle colliders, namely at the LHC.
Many LHC searches for DM  are based on the idea of looking at events with missing energy plus a single jet or photon, emitted from the initial state in $pp$ collisions
\begin{equation}
p p \to \chi+\overline{\chi}+{\rm jet},
\end{equation}
where $\chi$ indicates the DM particle. Several results are already available from two LHC collaborations \cite{monojetATLAS1, monojetCMS1, monojetATLAS2}. In order to avoid the overwhelming model-dependence introduced by the plethora of DM models discussed in the literature, DM searches at the LHC have made use of  the 
 Effective Field Theory (EFT) 
 \cite{Bai:2010hh,Goodman:2010ku}. However, as far as collider searches are concerned, with the LHC being such a powerful machine,
it is not guaranteed that the events used to constrain an effective interaction are not occurring
at an energy scale larger than the cutoff scale of the effective description. 
The question about the validity of the EFT for collider searches of DM has become
pressing (see also Refs.~\cite{Busoni:2013lha, Goodman:2010ku}), especially in the perspective of analysing the data from the future LHC run at (13-14) {\rm TeV}.

Let us consider a simple model where there is a heavy mediator of mass $M$, 
to which the quarks and DM are coupled
with couplings $g_q$ and $g_\chi$, respectively.
The EFT is a good approximation only at low energies. Indeed, it is possible at low energies to integrate out the heavy mediator from the theory and obtain a tower of operators.
The matching condition of the ultra-violet (UV) theory with the mediator and its low-energy effective counterpart implies $\Lambda={M}/{\sqrt{g_q g_\chi}}$.
A DM production event occurs at an energy at which the EFT is reliable as long as $Q_{\rm tr}<M$, where $Q_{\rm tr}$ is the  momentum transfer in the process;
this, together with the condition of perturbativity of the couplings $g_{q,\chi}<4\pi$, 
implies
\begin{equation}
\Lambda>\frac{Q_{\rm tr}}{\sqrt{g_qg_\chi}}>\frac{Q_{\rm tr}}{4\pi}\,.
\label{cond1}
\end{equation}
It is clear
that the details of condition (\ref{cond1}) depend on the values of the couplings in the UV
theory.
In the following, for definiteness, we will mostly identify the mass of the new degrees of freedom $M$ with
the suppression scale of the operator $\Lambda$. This is equivalent to consider couplings in the 
UV theory of ${\cal O}(1)$. So, we will deal with the condition (but we will discuss also  the impact of taking couplings larger than 1) 
\begin{equation}
Q_{\rm tr}\lesssim \Lambda\label{condition}\,.
\end{equation}
In Ref. \cite{Busoni:2013lha} we have started the discussion of
the limitations to the use of the EFT  approach for DM searches at the LHC by  adopting a toy model
where the heavy mediator is exchanged in the $s$-channel and by 
introducing  a few quantities which quantify the error made when using effective operators to describe processes with very high momentum transfer. Our criteria indicated up to what cutoff energy scale, and with what precision, the effective description is valid, depending on the DM mass and couplings.
%

\section{Validity of the EFT: analytical approach}
\label{sec:validity}

\subsection{Operators and cross sections}
The starting point of our analysis is the list  of the 18 operators listed in \cite{Busoni:2014sya} which are commonly used in the literature 
\cite{Goodman:2010ku}. 
We have considered not only the operators connecting the DM fermion to quarks (D1-D10), but also those involving gluon field strengths (D11-D14).
Furthermore, the operators can originate from heavy mediators exchange in the $s$-channel. 
For instance, 
 the D1' (D5) operators may be originated by the tree-level $s$-channel exchange of a very
heavy scalar (vector) boson
We have computed the tree-level differential cross sections in the transverse momentum $p_{\rm T}$ and rapidity $\eta$ of the final jet for the hard scattering process with gluon radiation from the initial state
$f+\bar f\to \chi+\bar \chi+g$, where $f$ is either a quark (for operators D1-D10),
or a gluon (for operators D11-D14).

In order to get the cross sections initiated by the colliding protons one needs to average
over the PDFs. 
We have performed the analytical calculation only for the  emission of an initial state gluon (identified with the final jet observed experimentally). The extension to include also the smaller contribution coming
from initial radiation of quarks ($qg\to\chi\chi+q$) is done numerically in Section \ref{sec:numerical}.

\subsection{Results and discussion}

\begin{figure}[t!]
\centering
\includegraphics[width=0.45\textwidth]{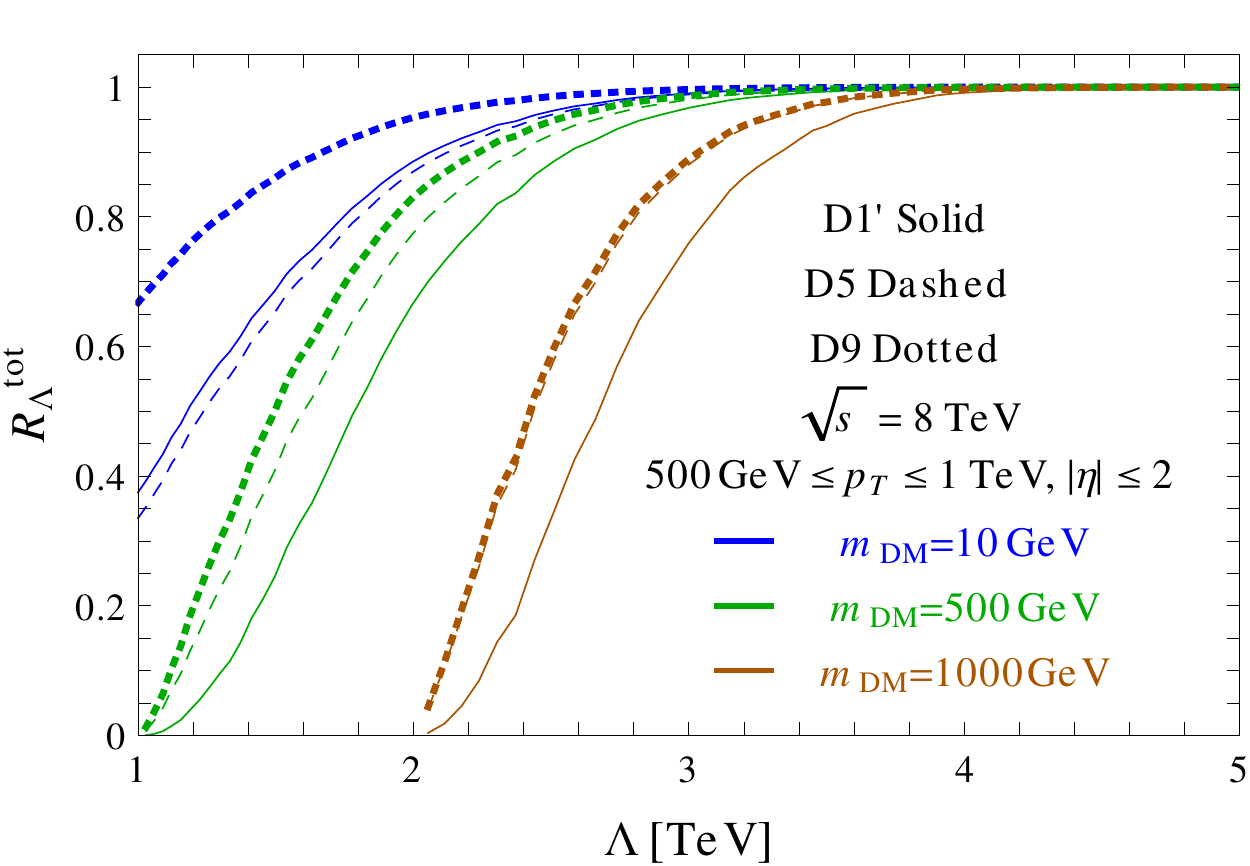}
\hspace{0.5cm}
\includegraphics[width=0.45\textwidth]{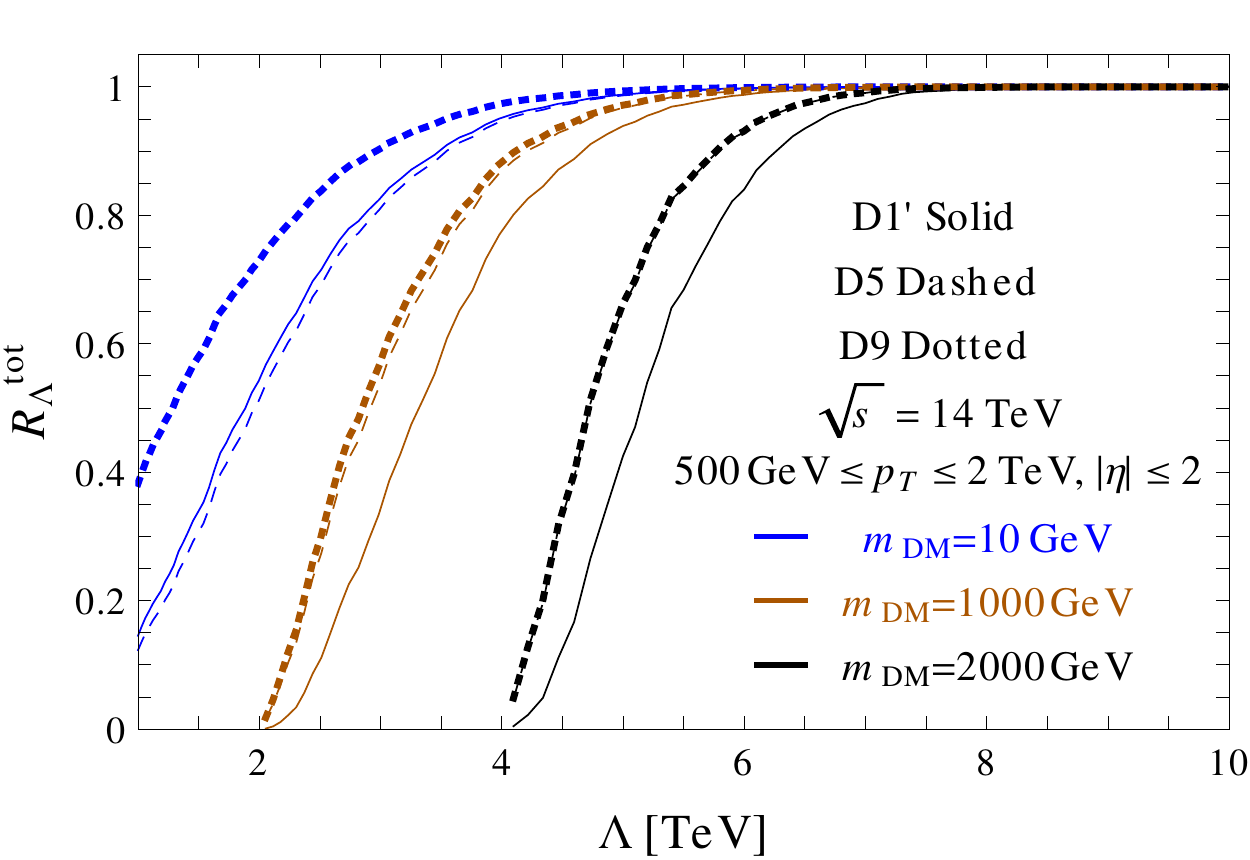}\\
\includegraphics[width=0.45\textwidth]{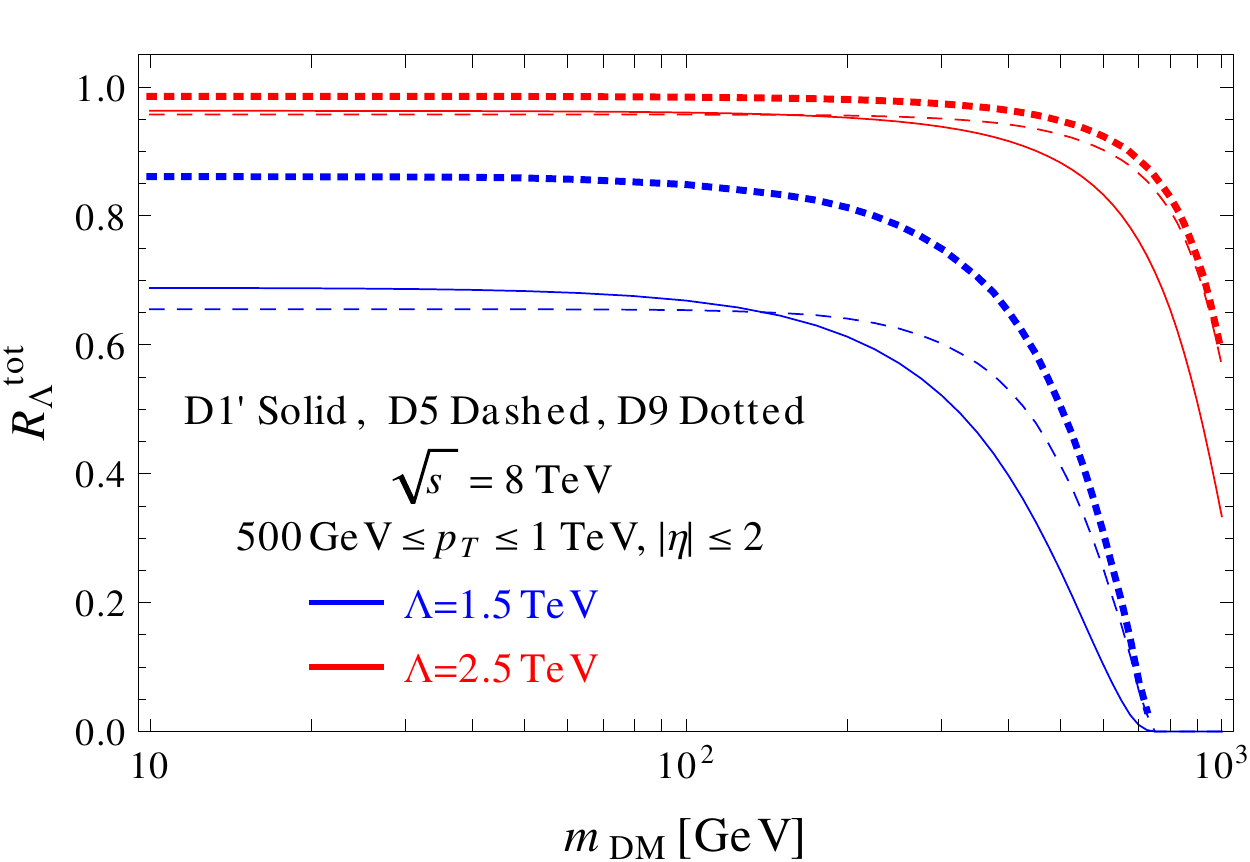}
\hspace{0.5cm}
\includegraphics[width=0.45\textwidth]{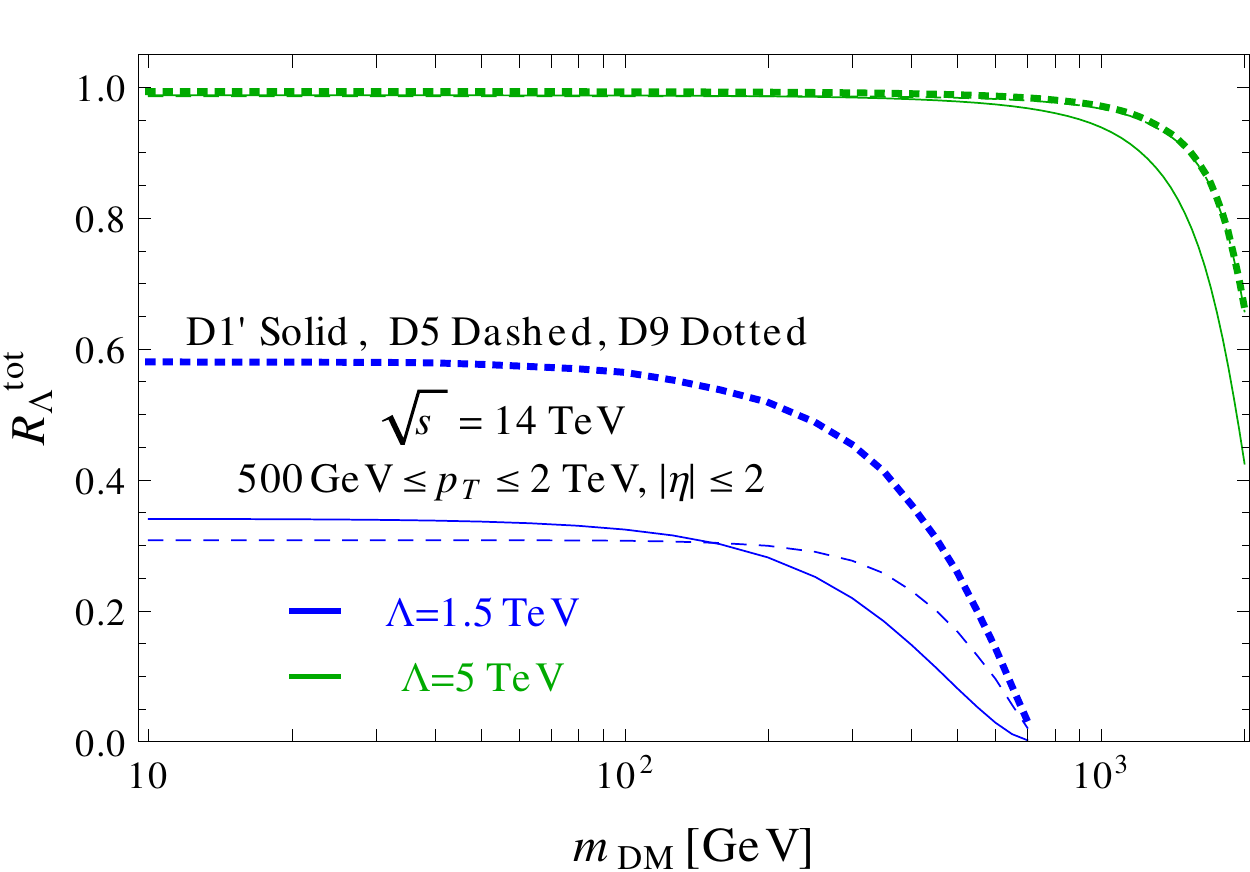}
\caption{ \em
 The ratio $R_{\Lambda}^{\rm tot}$ defined in Eq.(\protect\ref{ratiolambdatot}) for operators $D1'$ (solid lines),
  $D5$ (dashed lines) and $D9$ (dotted lines) as a function of $\Lambda$ and $m_{\rm DM}$,
  for $\sqrt{s}=8$ TeV (left panel) and $14$ TeV (right panel).}
\label{fig:RLambdatot}
\end{figure}

In what regions of the parameter space $(\Lambda, m_{\rm DM})$ is the effective description
accurate and reliable?
The truncation to the lowest-dimensional operator of the EFT expansion is  accurate only if the momentum transfer is smaller than an energy scale of the order of  $\Lambda$, see Eqs.(\ref{condition}). 
Therefore we want to compute the fraction of events with momentum transfer 
lower than the EFT cutoff scale.
To this end we define the ratio of  the  cross section
obtained in the EFT with the requirement $Q_{\rm tr}<\Lambda$ on the
PDF integration domain, over the total cross section obtained in the EFT.
\begin{equation}
R_\Lambda^{\rm tot}\equiv\frac{\sigma\vert_{Q_{\rm tr}<\Lambda}}{\sigma}
=\frac{\int_{p_{\rm T}^{\rm min}}^{p_{\rm T}^{\rm max}}\de p_{\rm T}\int_{-2}^2\de \eta
\left.\dfrac{\de^2\sigma}{\de p_{\rm T}\de\eta}\right\vert_{Q_{\rm tr}<\Lambda}}
{\int_{p_{\rm T}^{\rm min}}^{p_{\rm T}^{\rm max}}\de p_{\rm T}\int_{-2}^2\de \eta
\dfrac{\de^2\sigma}{\de p_{\rm T}\de\eta}}.
\label{ratiolambdatot}
\end{equation}
 The results are shown in Fig.\ref{fig:RLambdatot}. 
We show only results for representative operators $D1', D5, D9$.
This ratio $R_\Lambda^{\rm tot}$ gets closer to unity for large values of $\Lambda$, as in this case the effect of the cutoff becomes negligible. The ratio drops for large $m_{\rm DM}$ because the momentum
transfer increases in this regime.
This confirms our precedent  analysis of Ref.~\cite{Busoni:2013lha}, that
the EFT works better for large $\Lambda$ and small $m_{\rm DM}$.
Notice also that, going from $\sqrt{s}=8 {\rm TeV}$ to $\sqrt{s}=14 {\rm TeV}$, the results scale almost linearly with the energy, so for the same value of the ratio $m_{\rm DM}/\Lambda$ one obtains nearly the 
same $R_\Lambda^{\rm tot}$.

\begin{figure}[t!]
\centering
\includegraphics[width=0.45\textwidth]{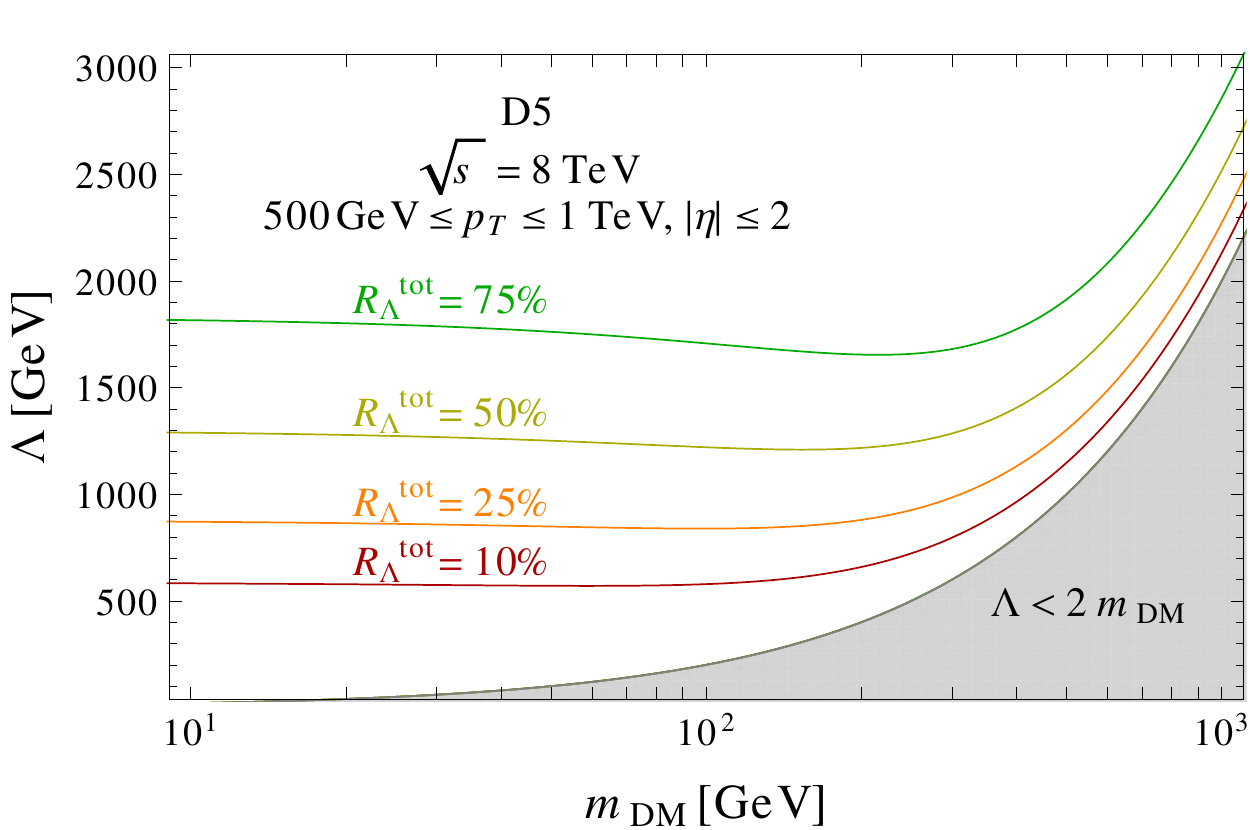}
\hspace{0.5cm}
\includegraphics[width=0.45\textwidth]{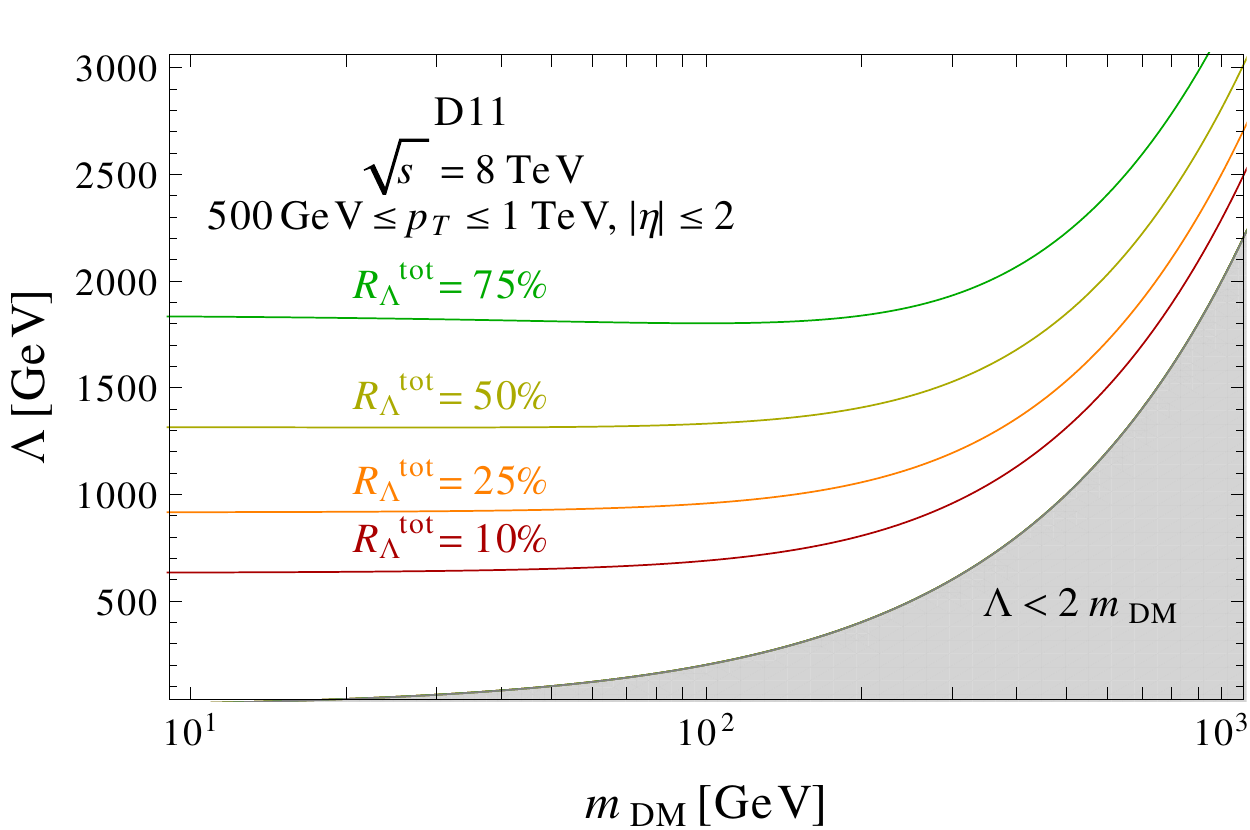}\\
\caption{ \em
Contours for the ratio $R_{\Lambda}^{\rm tot}$, defined in Eq.(\protect\ref{ratiolambdatot}), on the plane
$(m_{\rm DM}, \Lambda)$, for the different operators. We set $\sqrt{s}=8 {\rm TeV},
|\eta|\leq 2$ and $500 \GeV<p_{\rm T}<1 \TeV$. }
\label{fig:RLambdacontours}
\end{figure}

Next, we turn to study the contours of constant values of the quantity $R_{\Lambda}^{\rm tot}$,
in the plane $(m_{\rm DM}, \Lambda)$. These contour curves for the different operators
are shown in Fig.\ref{fig:RLambdacontours} for $\sqrt{s}=8$ {\rm TeV}.
The requirement that at least 50\% of the events occur with momentum transfer below the cutoff
scale $\Lambda$ requires such a cutoff scale to be above
$\sim 1 {\rm TeV}$ for $\sqrt{s}=8\, {\rm TeV}$, 
or above $\sim 2 {\rm TeV}$ for $\sqrt{s}=14\, {\rm TeV}$. 

To close this section let us comment on another 
 question one may ask: 
what is the difference between interpreting data with an effective operator and with
its simplest UV completion?
This question has  already been addressed in  Ref.~\cite{Busoni:2013lha} for the operator $D1'$, by
 studying the ratio of the cross sections obtained with the UV theory and with the effective operator.
 For each of the  operators listed in \cite{Busoni:2014sya} one can write a simple UV-complete Lagrangian.
The very same analysis can be repeated for all the other operators and we checked that
the same qualitative conclusions can be drawn.
In particular, if $\Lambda$ is not larger than a few {\rm TeV}, interpreting the experimental data in
terms of EFT or in terms of a simplified model with a mediator can make a significant difference.

\section{Comparison with MonteCarlo Simulations}
\label{sec:numerical}

In order to perform an alternative check of  our analytical results and  to be able to compare to the experimental limits as close as possible, we present 
in this section the results of numerical event simulations.

We made use of  \textsc{MadGraph 5}\cite{mg5}  to simulate $pp$ collisions at $\sqrt{s}=8$ {\rm TeV} and $\sqrt{s}=14$ {\rm TeV}. 
For details about this procedure, see \cite{Busoni:2014sya}.
According to the event kinematics we have  evaluated whether or not the conditions of validity discussed in Section \ref{sec:validity} are fulfilled. Specifically, we have  checked if Eqs.(\ref{cond1}) are fulfilled, that is, if the following condition is satisfied
\begin{equation}
\Lambda > \frac{Q_{\rm tr}}{\sqrt{g_q g_\chi}} > 2 \frac{m_{\rm DM}}{\sqrt{g_q g_\chi}}\, .
\label{kinconstraint}
\end{equation}

From the simulated samples the fraction of events fulfilling $\Lambda > Q_{\rm tr}/\sqrt{g_q g_\chi}$ for each pair of DM mass and cutoff scale can be evaluated, if one assumes a certain value for the couplings $\sqrt{g_\chi g_q}$.

\begin{figure}[t!]
\centering
\includegraphics[width=0.45\textwidth]{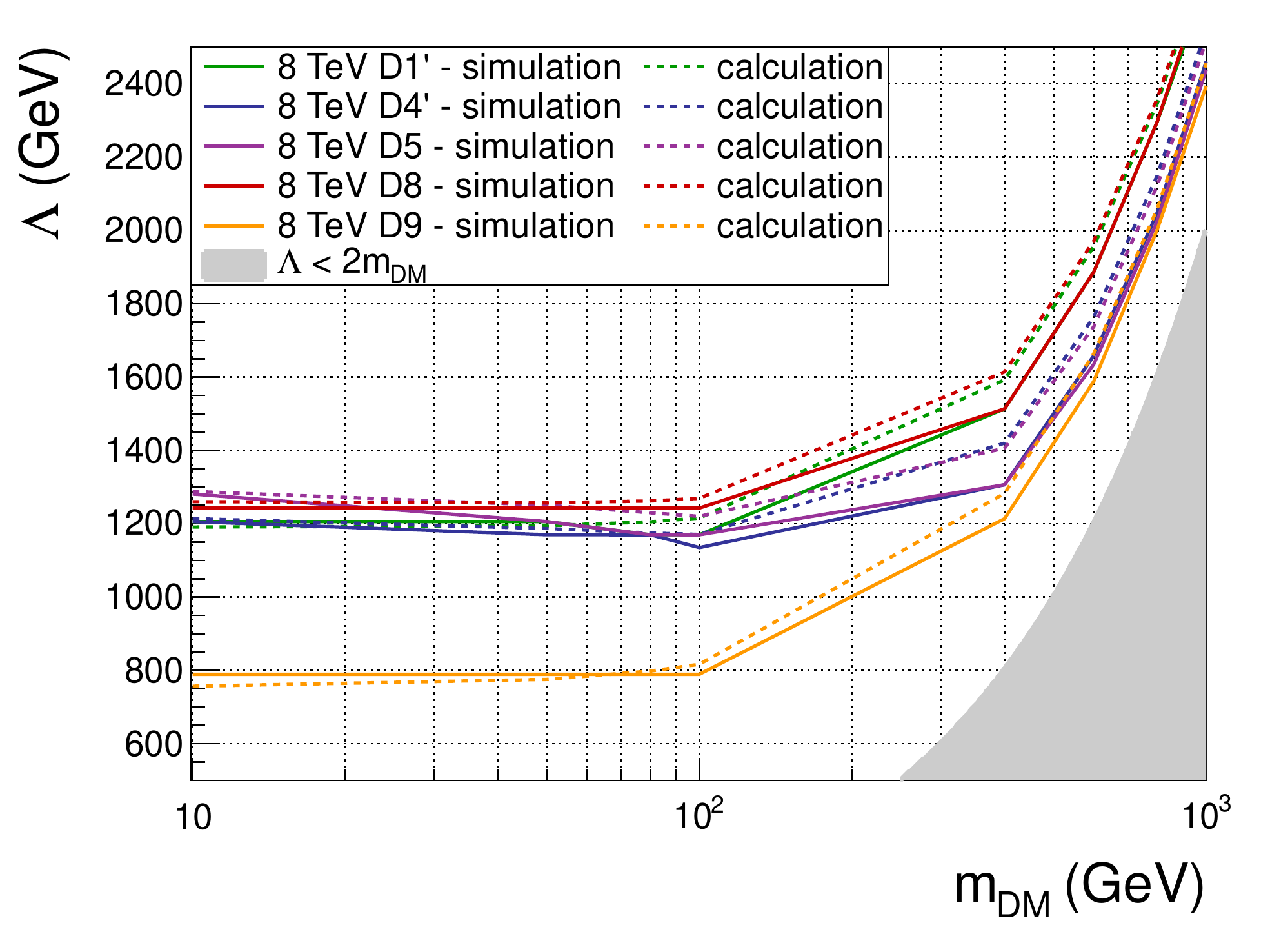}
\hspace{0.5cm}
\includegraphics[width=0.45\textwidth]{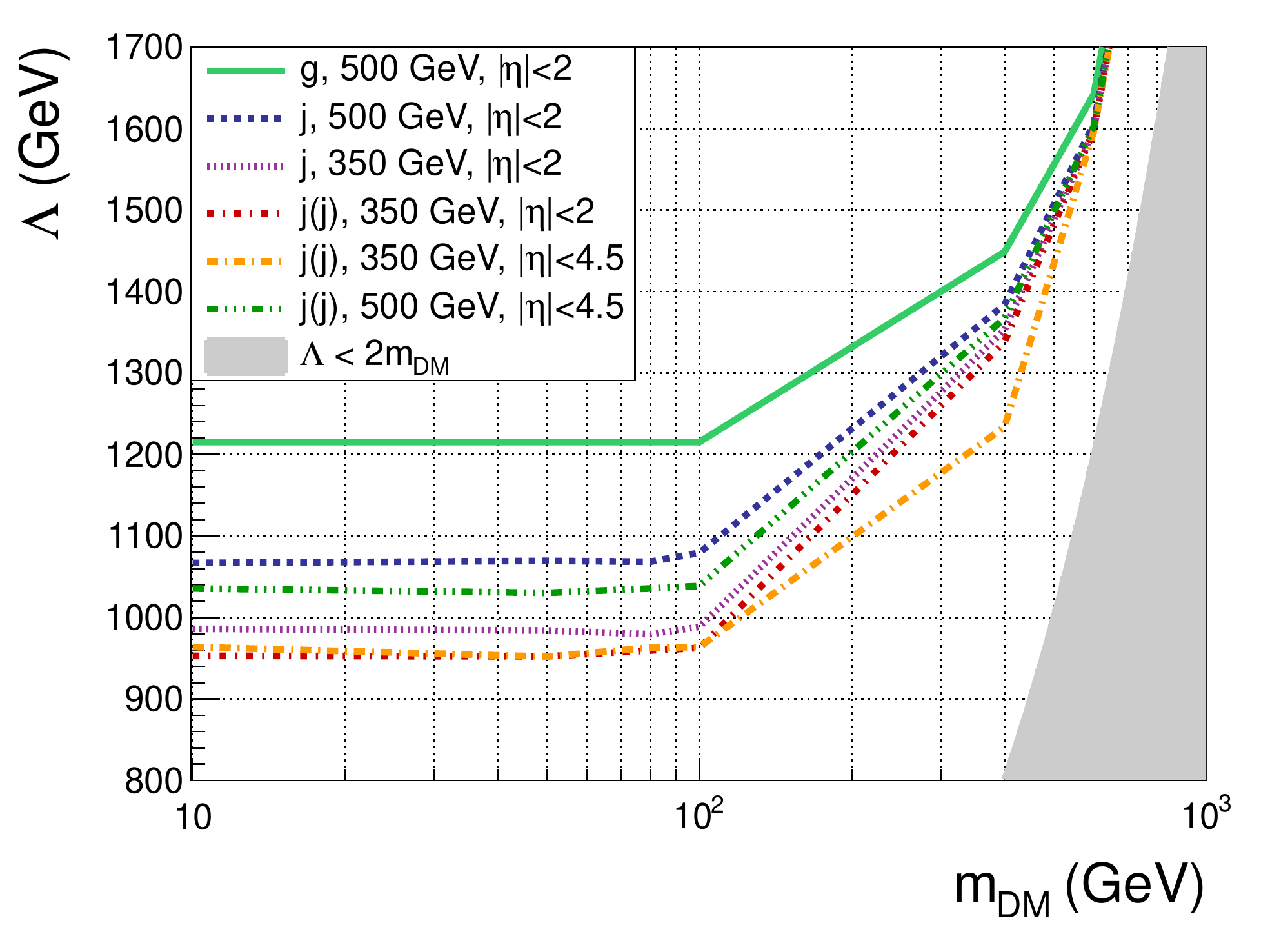}
\caption{ \em
(Left Panel)Comparison of the contour $R_\Lambda^{\rm tot} = 50 \%$ for the analytical calculation (dashed line) and the simulation (solid line) for the different operators $D1'$, $D4'$, $D5$, $D8$ and $D9$. The results agree within less than 7 \%. (Right Panel)  The changes of the contour of $R_\Lambda^{\rm tot} = 50 \%$ are shown for several variations from the analytically calculated scenario to a scenario close to the cuts used in the ATLAS monojet analysis exemplarily for the operator $D5$ at $\sqrt{s}=8$ TeV. In the legend, ``g'' means only gluon radiation, ``j'' stands for
either quark- or gluon-initiated jets, ``j(j)'' means a second jet is allowed.
}
\label{fig:calcsimcomp}
\end{figure}

In order to confirm that analytical and numerical results are in agreement, Figure~\ref{fig:calcsimcomp}  shows a comparison for the operators $D1'$, $D4'$, $D5$, $D8$ and $D9$. The contours of $R_\Lambda^{\rm tot} = 50 \%$ from analytical and numerical evaluation agree within less than 7 \%. The remaining differences could be due to the upper jet $p_{\rm T}$ cut not imposed during event simulation but needed for the analytical calculation, and the details of the fitting procedures. Next, we vary the kinematical constraints step by step from the scenario considered in the analytical calculations, to a scenario closest to the analysis cuts applied in the ATLAS monojet analysis~\cite{monojetATLAS2}. The effect of the variation of the cuts can be seen in Figure~\ref{fig:calcsimcomp}.
Moving to the scenario closer to the experimental analysis leads to contours that are at most  $\sim30\%$ lower in $\Lambda$. After having extracted $R_\Lambda^{\rm tot}$ for each WIMP and mediator mass, a curve can be fitted through the points obtained in the plane of $R_\Lambda^{\rm tot}$ and $\Lambda$. See \cite{Busoni:2014sya} for more details.


\section{Implications of the limited validity of EFT in DM searches at LHC}
\label{sec:interp}

\begin{figure}[t!]
\centering
\includegraphics[width=0.45\textwidth]{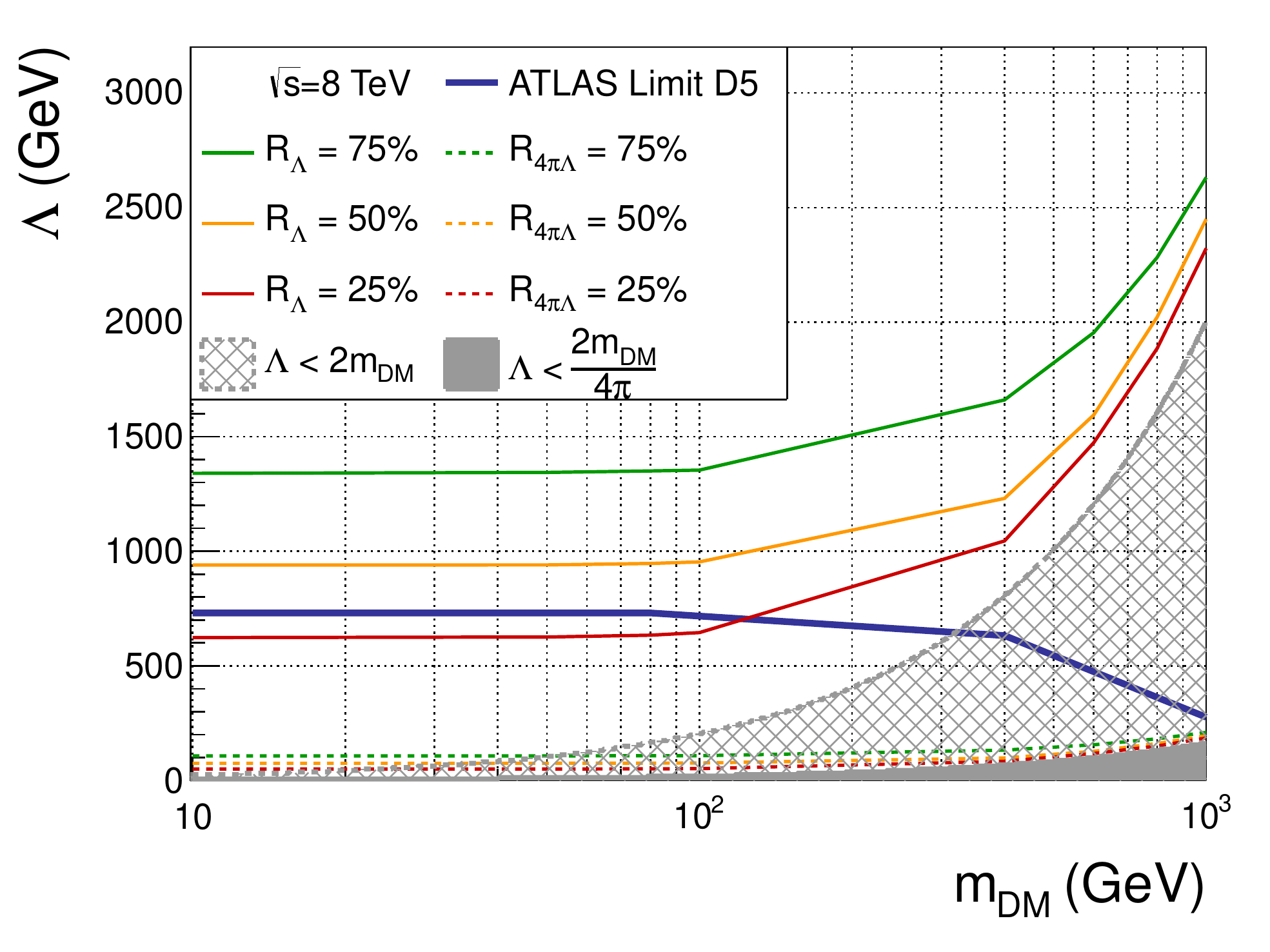}
\hspace{0.5cm}
\includegraphics[width=0.45\textwidth]{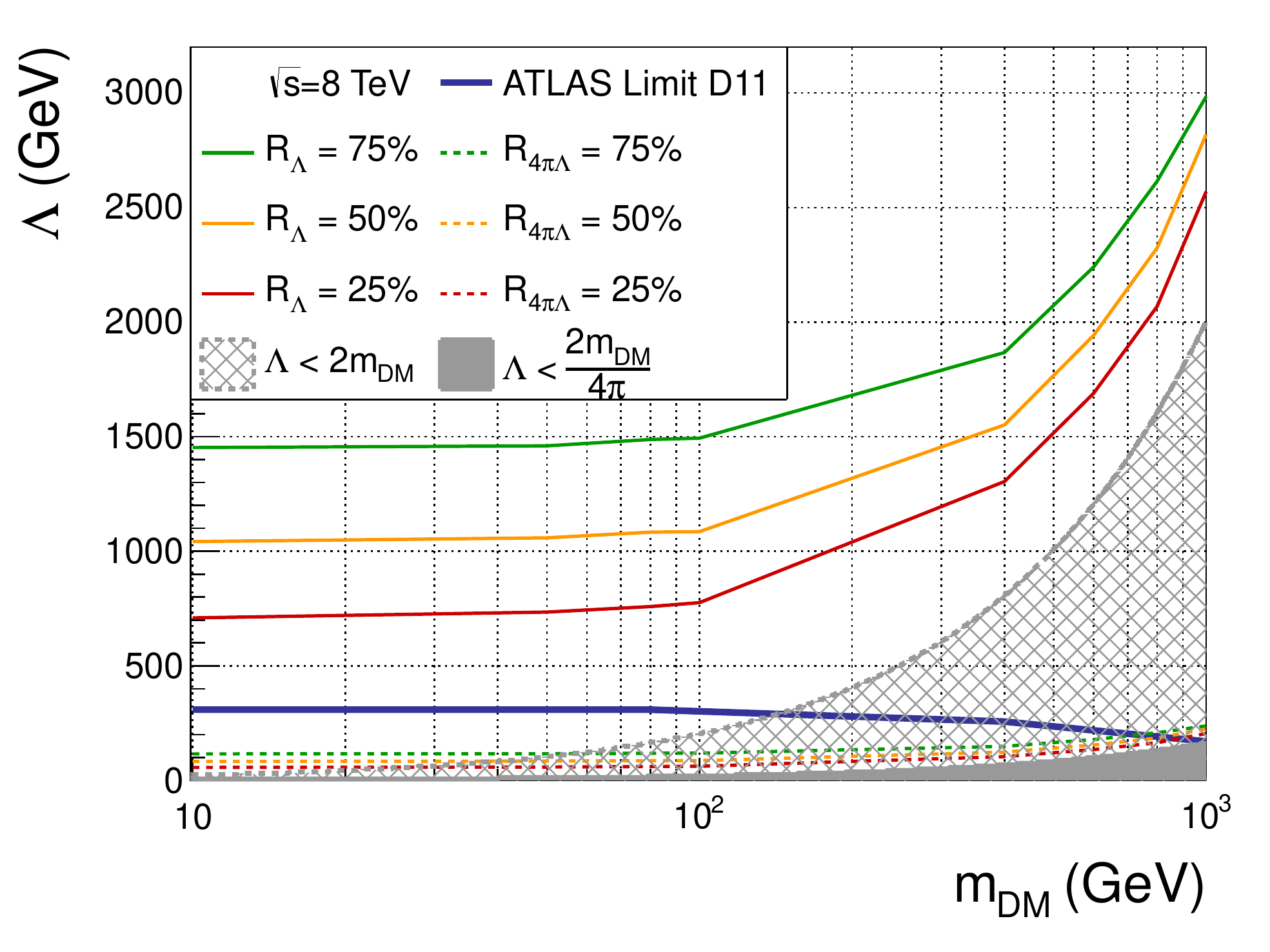}
\caption{\em
25\%, 50\% and 75\% contours for the ratio $R_{\Lambda}^{\rm tot}$, compared to the experimental limits from ATLAS~\cite{monojetATLAS2} (blue line). Also indicated are the contours of $R_{\Lambda}^{\rm tot}$ in the extreme case when setting the couplings $\sqrt{g_q g_\chi}=4\pi$ (dashed lines). Results are shown for different operators: D5 (left panel) and D11 (right panel).
 }
\label{fig:expcomp}
\end{figure}

Figure \ref{fig:expcomp} shows the experimental limits obtained from the ATLAS monojet  analysis
 \cite{monojetATLAS2} in the plane $(\Lambda$, $m_{\rm DM}$), for the opearators D5 and D11.
The contours of $R_{\Lambda}^{\rm tot}$ for 25\%, 50\% and 75\% are superimposed. The experimental limits are placed in a region where only about 30\% of the events can be expected to fulfill the EFT conditions. Especially the limit on the gluon operator $D11$ seems questionnable. For comparison, dashed lines show the contours of $R_{\Lambda}^{\rm tot}$ for the case of $\sqrt{g_q g_\chi}=4\pi$, presenting the limiting case for which the theory is still considered perturbative.

\begin{figure}[t!]
\centering
\includegraphics[width=0.45\textwidth]{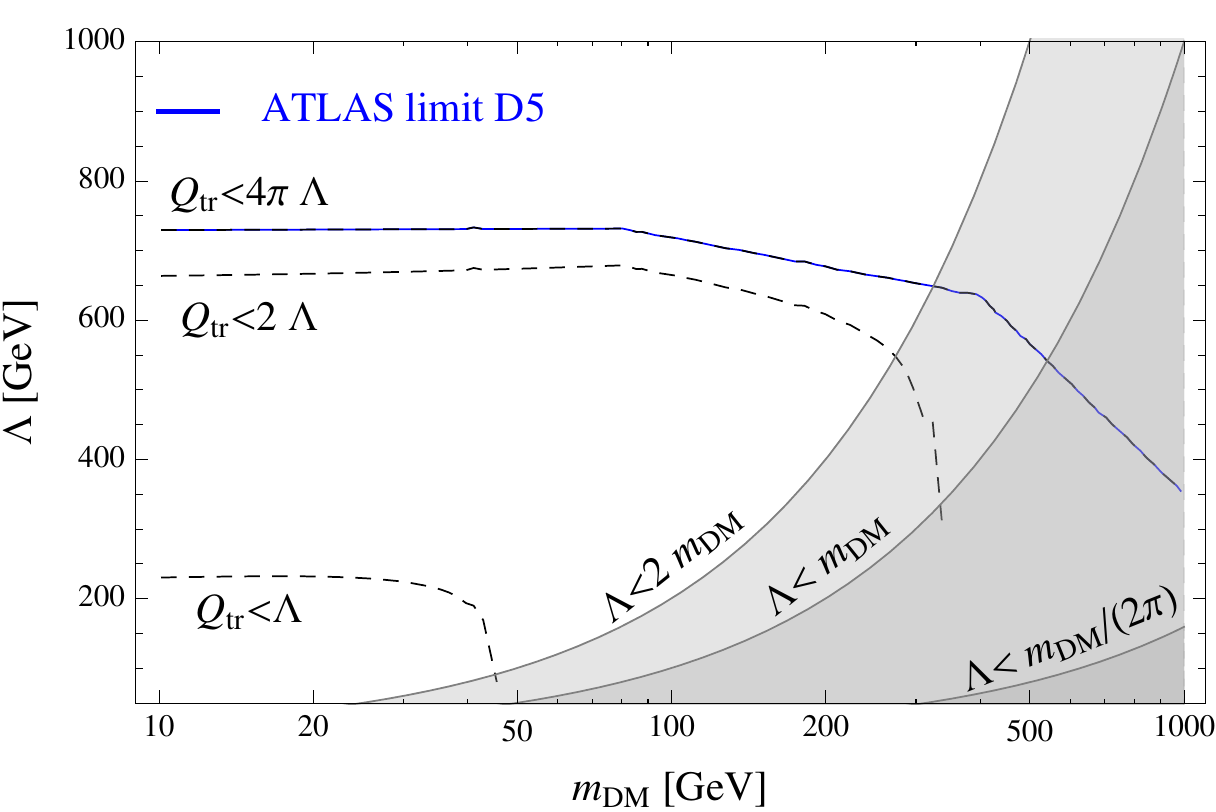}
\hspace{0.5cm}
\includegraphics[width=0.45\textwidth]{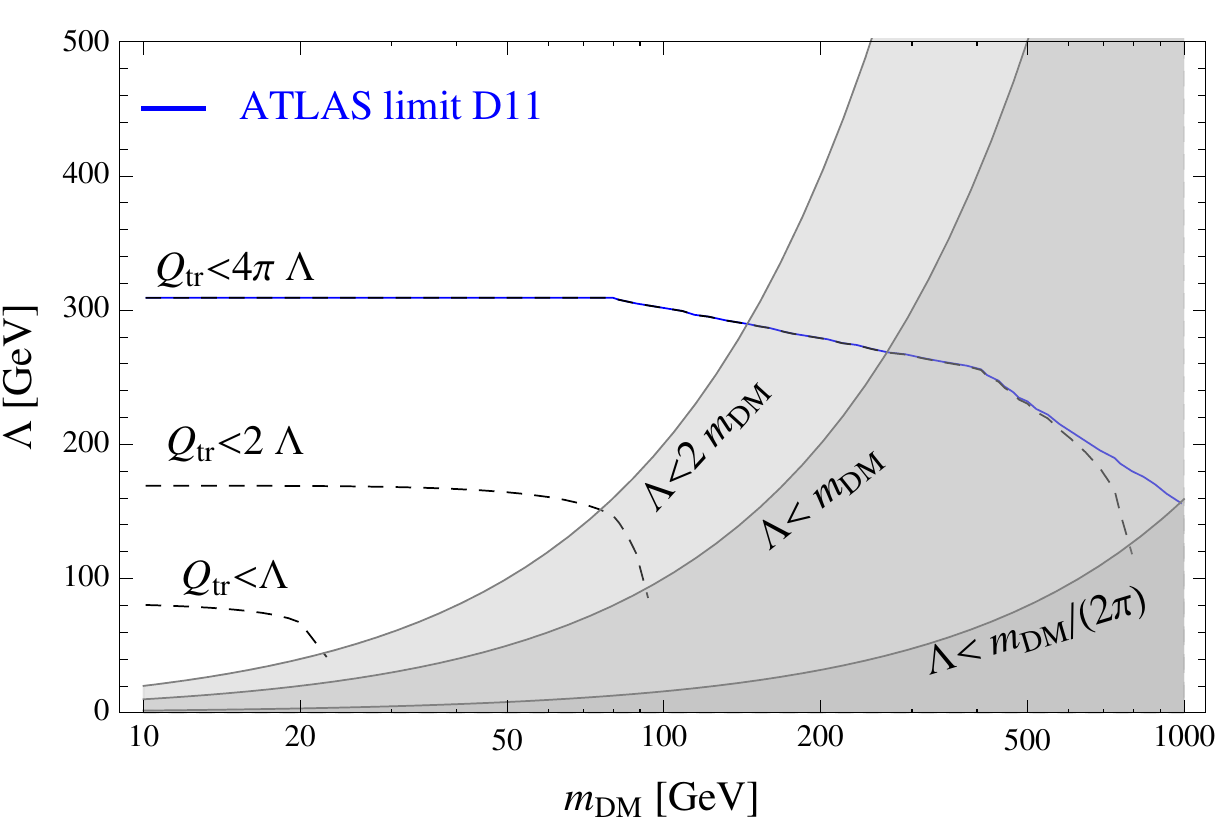}
\caption{ \em
The experimental limits by ATLAS \cite{monojetATLAS2} on the suppression scale $\Lambda$
are shown as solid blue lines.
The updated limits  taking into account EFT validity are shown as dashed black lines,
for $Q_{\rm tr}<\Lambda, 2\Lambda, 4\pi\Lambda$,
corresponding to different choices of the UV couplings: $\sqrt{g_qg_\chi}=1, 2, 4\pi$, respectively.
The corresponding kinematical constraints (Eq.(\protect\ref{kinconstraint})) are denoted by gray bands.
The different plots refer to different operators: D5  (left panel) and D11 (right panel). }
\label{fig:newlimit}
\end{figure}

In Fig.\ref{fig:newlimit} we show the new limits for the operators D5 and D11, for the conditions $Q_{\rm tr}<\Lambda, 2\Lambda, 4\pi\Lambda$,
corresponding different choices of the UV couplings: $\sqrt{g_qg_\chi}=1, 2, 4\pi$, respectively.
The  ATLAS bound reported is the 90\%CL observed limit.
The functions $R_\Lambda^{\rm tot}$ used are taken from the fitting functions described in \cite{Busoni:2014sya}, which include both quark and gluon jets, and the same cuts as the ``Signal Region 3''
used by ATLAS.
As expected, the weaker is the condition on $Q_{\rm tr}$, the more the new limits approach the ATLAS bound.
In the case of extreme couplings $\sqrt{g_qg_\chi}=4\pi$, while for D5 the new limit is indistinguishable
from the ATLAS one, for D11 the bound at large DM masses still need to be corrected.
In general, for couplings of order one, the limits which are safe from the EFT point of view are appreciably
weaker than those reported. We encourage the experimental collaborations to take this point into account when publishing their limits.

\section{Conclusions}
\label{sec:conclusions}
Following Ref. \cite{Busoni:2013lha}, we have studied the quantity $R_\Lambda^{\rm tot}$
(see Eq.(\ref{ratiolambdatot}), which quantifies the error made when using EFT to describe processes with very high momentum transfer. Our criterion indicates up to what cutoff energy scale the EFT is valid, depending on the DM mass and couplings.
We have performed the analysis for the full list of EFT operators,  connecting fermion DM particles and quarks or gluons,  used by the ATLAS and CMS collaborations and originated from the exchange of heavy mediators in the $s$-channel. 
We have also extended our analysis to the case of $\sqrt{s}=$ 14 {\rm TeV}. Furthermore,
we have validated our analytical results by performing  numerical event simulations which reproduce the experimental situation 
in the  closest possible  way. Our results indicate that the range of validity of the EFT is significantly limited in the parameter space
$(\Lambda,m_{\rm DM})$. While our findings are valid for the $s$-channel, a similar analysis is under way for the $t$-channel \cite{Busoni:2014haa} where similar results are obtained.

Does it mean that the EFT is not the best tool to interpret the current LHC data of DM searches? The answer is yes and no. On the negative  side, our results clearly cry out for an overcoming of the EFT, most possibly through 
identifying  a handful of classes of models (able to reproduce the EFT operators in the heavy mediator limit); this would allow a consistent
analysis of the current and future LHC data by consistently taking into account the role played by the mediator. On the positive side, keep working with the EFT allows to avoid the overwhelming model-dependence generated by the many DM models proposed so far. 
Nonetheless,  as we have shown in section \ref{sec:interp}, the price to pay 
is  a deterioration of the limits presented so far.


\begin{thebibliography}{99}

\bibitem{Busoni:2014sya}
  G.~Busoni, A.~De Simone, J.~Gramling, E.~Morgante and A.~Riotto,
  \href{http://arXiv.org/abs/1402.1275}{[arXiv:1402.1275]} [hep-ph].

\bibitem{monojetATLAS1}
G.~Aad {\it et al.}  [ATLAS Collaboration],
  JHEP {\bf 1304}, 075 (2013) 
  \href{http://arXiv.org/abs/1210.4491}{[arXiv:1210.4491]}.

\bibitem{monojetCMS1}
  S.~Chatrchyan {\it et al.}  [CMS Collaboration],
  Phys.\ Rev.\ Lett.\  {\bf 107}, 201804 (2011)
\href{http://arXiv.org/abs/1106.4775}{[arXiv:1106.4775]}.  
  
\bibitem{monojetATLAS2}
\href{http://cds.cern.ch/record/1493486}{ATLAS-CONF-2012-147}.	 


\bibitem{Bai:2010hh}  
Y. Bai, P. J. Fox and R. Harnik, JHEP {\bf 1012}, 048 (2010)  \href{http://arXiv.org/abs/1005.3797}{[arXiv:1005.3797]}. 
  
\bibitem{Goodman:2010ku}    
J.~Goodman, M.~Ibe, A.~Rajaraman, W.~Shepherd, T.~M.~P.~Tait and H.~-B.~Yu,
  Phys.\ Rev.\ D {\bf 82}, 116010 (2010)
    	  \href{http://arXiv.org/abs/1008.1783}{[arXiv:1008.1783]}. 

\bibitem{Busoni:2013lha}
  G.~Busoni, A.~De Simone, E.~Morgante and A.~Riotto,
  Physics Letters B 728C (2014)
\href{http://arXiv.org/abs/1307.2253}{[arXiv:1307.2253]}.  
     	  
\bibitem{mg5}
J.~Alwall, M.~Herquet, F.~Maltoni, O.~Mattelaer, T.~Stelzer,
JHEP 1106(2011)128 \href{http://arxiv.org/abs/1106.0522}{[arXiv:1106.0522]}.

\bibitem{Busoni:2014haa}
  G.~Busoni, A.~De Simone, T.~Jacques, E.~Morgante and A.~Riotto,
  \href{http://arXiv.org/abs/1405.3101}{[arXiv:1405.3101]} [hep-ph].
  
\end{thebibliography}
\end{document}